\documentclass[dvipsnames]{article}
\usepackage{fullpage}
\usepackage[square,numbers]{natbib}
\usepackage{color}
\usepackage{amsmath}
\usepackage{xcolor}
\usepackage{hyperref}
\usepackage[inline, shortlabels]{enumitem}
\usepackage{algorithm}
\usepackage{algpseudocode} 
\usepackage{graphicx}
\usepackage{subcaption}
\usepackage{booktabs}

\setlist{itemjoin ={,\enspace},itemjoin* = { and\enspace}}

\newcommand{\frSet}[1]{\mathcal{#1}}

\newcommand{\frDocument}[0]{d}
\newcommand{\frDocuments}[0]{\frSet{D}}

\newcommand{\frQueries}[0]{\frSet{Q}}
\newcommand{\frQuery}[0]{q}
\newcommand{\frAuthor}[0]{a}
\newcommand{\frAuthors}[0]{\frSet{A}}
\newcommand{\frExposure}[0]{e}
\newcommand{\frGroupExposure}[0]{\mathcal{E}}

\newcommand{\frEE}[0]{\Delta_\frGroups}

\newcommand{\frRanking}[0]{\pi}
\newcommand{\frRankings}[0]{\Pi}

\newcommand{\frRelevance}[0]{r}

\newcommand{\frGroups}[0]{\frSet{G}}
\newcommand{\frGroup}[0]{g}
\newcommand{\frNumGroups}[0]{|\frGroups|}

\newcommand{\frRelevanceTransform}[0]{f}

\begin{document}
\title{Overview of the TREC 2020 Fair Ranking Track\footnote{Data and code are available at: \url{https://fair-trec.github.io/2020/}}}
\author{
  Asia J. Biega\\
  Microsoft Research Montréal\\
  \texttt{asia.biega@acm.org}\\
  \\
  Michael D. Ekstrand\\
  Boise State University\\
  \texttt{michaelekstrand@boisestate.edu}\\
  \\
    Sebastian Kohlmeier\\
  Allen Institute for Artificial Intelligence\\
  \texttt{sebastiank@allenai.org}
  \and
  Fernando Diaz\\
  Montreal Institute for Learning Algorithms\\
  \texttt{diazf@acm.org}\\
  \\
  Sergey Feldman\\
  Allen Institute for Artificial Intelligence\\
  \texttt{sergey@allenai.org}
}
\date{\empty}
\maketitle

For 2020, we again adopted an academic search task, where we have a corpus of academic article abstracts and queries submitted to a production academic search engine.   The central goal of the Fair Ranking track is to provide \emph{fair exposure} to different groups of authors (a \emph{group fairness} framing).  We recognize that there may be multiple group definitions (e.g. based on demographics, stature, topic) and hoped for the systems to be robust to these.  We expected participants to develop systems that optimize for fairness and relevance for arbitrary group definitions, and did not reveal the exact group definitions until \emph{after} the evaluation runs were submitted. 

The track contains two tasks, \emph{reranking} and \emph{retrieval}, with a shared evaluation.

\begin{description}
    \item[Rerank] runs sorted a query-dependent list of documents to simultaneously provide fairness and relevance.
    \item[Retrieval] runs returned 100-item rankings from the corpus in response to a query string.
\end{description}

The track organizers provided a sequence of queries, each accompanied by a varying-size set of documents.
Both tasks used the same queries; participants were asked not to use the test queries' rerank sets as a component of their retrieval model training.

\section{Protocol}
For our fair ranking evaluation, we provided participants with a sequence $\frQueries$ of queries accompanied by unordered sets of documents to rank.
The document sets are of varying size. For each request (query $\frQuery$ and set of documents $\frDocuments_\frQuery$), participants provided a ranked list of the documents from $\frDocuments_\frQuery$. 
For the retrieval task, $\frQuery$ is the set of all documents in our corpus and participants were asked to return a fixed set of documents.  
The final system output is a sequence of rankings for each query. Algorithm~\ref{protocol} presents a pseudocode of the evaluation protocol.

The rankings produced in response to queries in the sequence were to balance two goals:
\begin {enumerate*} [(1) ]%
\item be relevant to the consumers
\item be fair to the producers.
\end {enumerate*}

\begin{algorithm}
\caption{Evaluation protocol}\label{protocol}
\begin{algorithmic}
\State $\forall \frQuery,\frDocuments_\frQuery\in\frQueries, \frRankings_{\frQuery}\gets \{\}$
\For{$\frQuery,\frDocuments_\frQuery\in\frQueries$}
\State $\frRanking\gets \Call{\textcolor{red}{System}}{\frQuery, \frDocuments_\frQuery}$
\State $\frRankings_\frQuery\gets \frRankings_\frQuery \cup \{\frRanking\}$
\EndFor
\State \Return $\left\{\frRankings_\frQuery\right\}$
\end{algorithmic}
\end{algorithm}

\section{Evaluation}
\label{sec:evaluation}
Unlike previous TREC tracks, systems were to return multiple rankings for each query, as they might in response to different impressions of the same query text.
At evaluation time, we measured measure \emph{expected exposure} of groups over rankings produced for each given query \cite{diaz:expexp}.

Given a sequence of queries $\frQueries$ and associated system rankings, we evaluated systems according to fair exposure of authors and relevance of documents.

\subsection{Measuring Author Exposure for a Single Query}
\label{sec:attention}
\begin{figure}
	\begin{center}
	\includegraphics[width=2in]{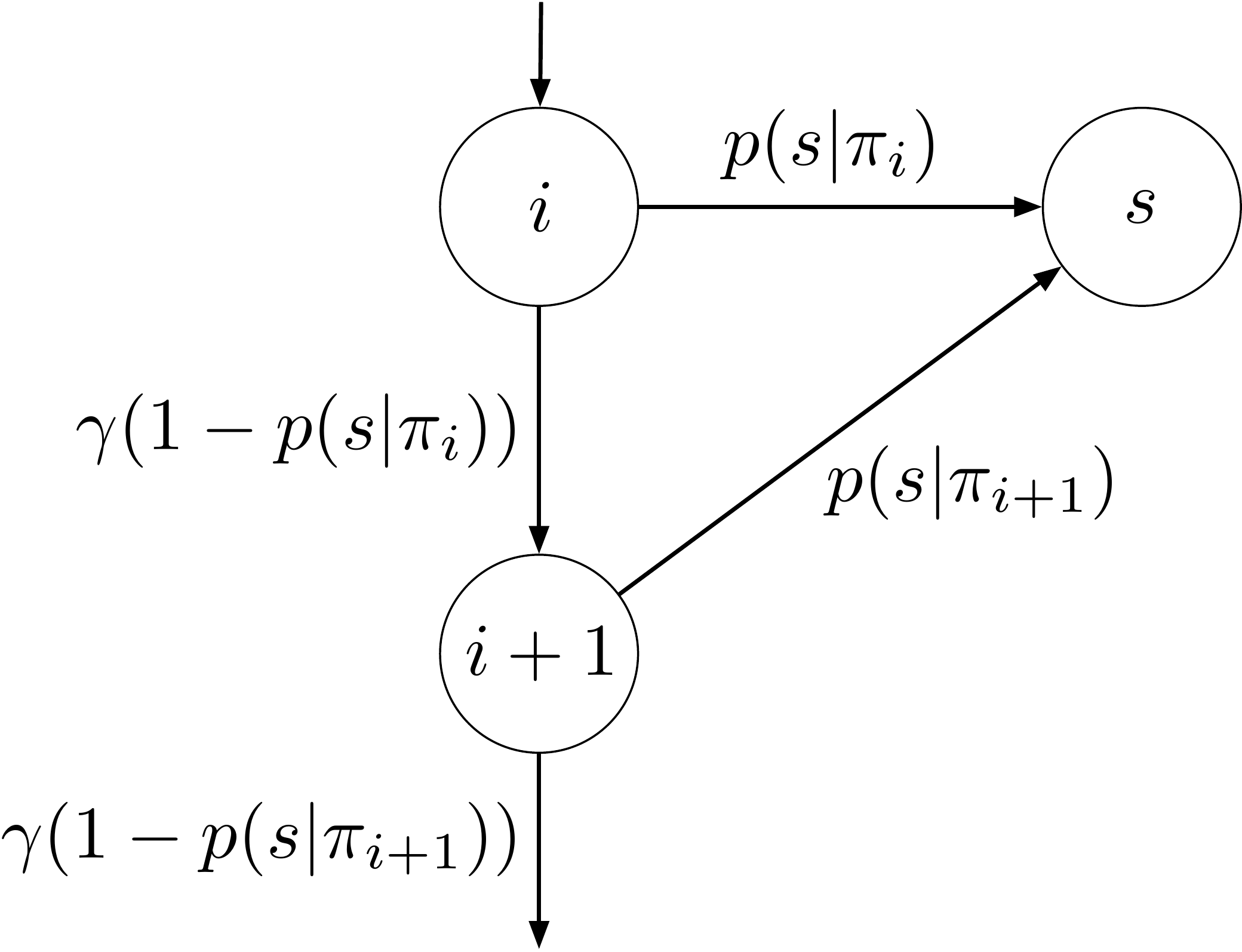}
	\end{center}
	\caption{Attention model.}
	\label{fig:attention}
\end{figure}

In order to measure exposure, we adopt the browsing model underlying the Expected Reciprocal Rank metric \cite{chapelle:err}.  Given a static ranking $\frRanking$ in response to a query impression, the exposure of author $\frAuthor$ is,
\begin{align*}
\frExposure_\frAuthor^\frRanking &= \sum_{i=1}^n \left[\gamma^{i-1}\prod_{j=1}^{i-1}(1-p(s|\pi_j))\right] I(\pi_i\in\frDocuments_\frAuthor)\\
\\
n & \phantom{=} \text{number of documents in ranking }\pi\\
\frDocuments_\frAuthor & \phantom{=} \text{documents including $\frAuthor$ as an author}\\
\pi_i & \phantom{=} \text{document at position $i$}\\
\gamma & \phantom{=} \text{continuation probability (fixed to 0 for the final position in the ranking)}\\
p(s|\frDocument) & \phantom{=} \text{probability of stopping given user examined $\frDocument$}
\end{align*}
We present a graphical depiction of this model in Figure \ref{fig:attention}.

We used a discounting factor $\gamma=0.5$,
and assumed $p(s|\frDocument) = \frRelevanceTransform(\frRelevance_\frDocument)$, where $\frRelevance_\frDocument$ is the relevance of the document $\frDocument$ and $\frRelevanceTransform$ is a monotonic transform of that relevance into a probability of being satisfied.  

In order to compute the \emph{expected exposure} for $\frAuthor$, we consider the set of all rankings presented by the system for that query $\frRankings_\frQuery$,
\begin{align}
	\frExposure_\frAuthor &= \sum_{\frRanking\in\frRankings_\frQuery} \frExposure_\frAuthor^\frRanking \label{eq:ee}
\end{align}

The \emph{target expected exposure} for a query is derived from Equation \ref{eq:ee} assuming a policy that randomizes amongst all permutations whose relevance monotonically degrades with rank \cite{diaz:expexp}.

\subsection{Measuring Group Exposure for a Single Query}
Assume that each author is assigned to exactly one of $\frNumGroups$ groups.  Let $\frAuthors_\frGroup$ be the set of all authors in group $\frGroup$.  The group expected exposure is defined as,
\begin{align}	
	\frGroupExposure_\frGroup &= \sum_{\frAuthor \in \frAuthors_\frGroup} \frExposure_\frAuthor \label{eq:gexpexp}
\end{align}
We define the target group expected exposure $\frGroupExposure^*_\frGroup$ as Equation \ref{eq:gexpexp} using the individual target expected exposure (Section \ref{sec:attention}).
\subsection{Expected Exposure Metric}
We evaluated systems using the per-query difference in system group expected exposure and target group expected exposure,
\begin{align}
	\frEE(\frGroupExposure)&=\left(\sum_{\frGroup\in\frGroups}(\frGroupExposure_\frGroup - \frGroupExposure^*_\frGroup)^2\right)^{\frac{1}{2}}\label{eq:eemetric}
\end{align}
We averaged per-query metrics to compute the summary metric for the run.  
\section{Data}

\label{sec:data}

\subsection{Input}

Three main inputs were made available to participants: the \emph{corpus} of articles to search, the \emph{example group definition} file
to help them develop and test their solutions, and the \emph{queries}.

\subsubsection{Paper and Author Data}

The paper and author metadata CSV files provide summary information for papers and their authors in the rerank set.
There are three files:

\begin{description}
    \item[paper\_metadata.csv] contains basic paper information: ID, title, year, venue, and the number of citations.
    \item[author\_metadata.csv] contains author information: ID, name, citation count, paper count, and H-index.
    \item[authors\_for\_papers.csv] contains the author list for each paper: paper ID, author ID, and position.
\end{description}

These files do \emph{not} contain abstracts.  Creating a usable index requires the corpus in the next section.

\subsubsection{Corpus}

The full corpus for this track was the Semantic Scholar (S2) Open Corpus from the Allen Institute for Artificial Intelligence.
It can be downloaded from \url{http://api.semanticscholar.org/corpus/}, and consists of 186 1GB data files.
Each file is compressed JSON, where each line is a JSON object describing one paper.
The following data are available for most papers:

\begin{itemize}
    \item S2 Paper ID
    \item DOI
    \item Title
    \item Abstract
    \item Authors (resolved to author IDs)
    \item Inbound and outbound citations (resolved to S2 paper IDs)
\end{itemize}

We provide tools for subsetting the corpus at \url{https://github.com/fair-trec/fair-trec-tools}.
These tools were used to create the subset we released to participants.

\subsubsection{Example Group Definition Data}

For training, we provided the file \texttt{fair-TREC-sample-author-groups.csv} containing group ids for authors in the S2
corpus. 
This group definition was not our final group definition, but was intended to help groups get started on the task.

This CSV file contains two columns:

\begin{enumerate}
    \item The \texttt{author} column has the S2 ID of the author.
    \item The \texttt{gid} column has the author's group identifier.
\end{enumerate}

\subsubsection{Queries}
\label{sec:queries}

\paragraph{Query data.}
The query data was obtained from searches that occurred on the Semantic Scholar\footnote{\url{https://www.semanticscholar.org}} website between Feb 14, 2020 and April 27, 2020. The data consisted of session id, query text, result papers from the first 3 pages (30 results), and result clicks. 
Sessions with more than 25 unique queries were excluded, after which only sessions with at least 1 result paper click and no more than 250 result paper clicks were included. 

\paragraph{Query-document relevance.}
We estimated the relevance of different documents to queries based on the click data described above. We computed the query-document relevance as a weighted average of the number of clicks on a given document over all impressions of a given query-document pairs present in the data. For weighting, we used ranking position propensity scores estimated by the Semantic Scholar from their system data. Relevance scores were converted to binary based on a manually selected threshold.

\paragraph{Query filtering.}
Because of the exhaustive annotation process that required annotating group memberships of all document authors, we then sampled a smaller number of queries to construct evaluation sequences. We released $200$ training and $200$ evaluation queries. For both the training and evaluation data, these queries were selected first by random sampling, and then by a number of filtering steps. More specifically,
\begin{itemize}
    \item To help remove known-item queries, we included only queries with at least two relevant documents and excluded queries with more than 4 words.
    \item We further manually cleaned the sample to remove any known-item queries, queries containing people's names, and queries with offensive and sensitive keywords.
\end{itemize}

\paragraph{Query sequences.}
Since the evaluation this year focused on individual queries, each query sequence consisted of repetitions of a single query. We had 200 sequences, each consisting of a 100 repetitions of a query.

\subsection{Output}

For each query sequence, participants submitted a JSON file where each line is a JSON object (a dictionary) containing their ranking results:

\begin{itemize}
    \item $<$sequence id$>$.$<$query number in sequence$>$ (`q\_num')
    \item $<$query id$>$ (to look up in query file) (`qid')
    \item An ordered list of document IDs (of the documents to be re-ranked for the query) (`ranking')
\end{itemize}

\subsection{Annotations}

NIST assessors annotated returned papers with the country in which each author was operating (based on their affiliation data in the paper manuscript), along with institution type (academic, industry, nonprofit, government, etc.). 
Not all papers were able to be annotated.  These are the known reasons a paper may not have annotations:

\begin{itemize}
    \item It has a large author list ($> 10$). We excluded such long papers because there were not very many of them, and large-team papers require special treatment in how we consider their author lists, particularly when authors may be from different groups.
    \item Some papers did not have an accessible source with sufficient affiliation information to provide annotations (e.g. no available PDF file, and a paper information page that either did not contain affiliation details or was not accessible from the annotation interface).
    \item Some papers may not provide sufficient information to determine an author's affiliation location.
\end{itemize}

All documents in the candidate sets for the rerank tests were annotated, along with many of the documents in the  corpus for the retrieval task.

\begin{table}[tbh]
    \centering
    \begin{tabular}{lrr}
{} &  Overall &  Eval Candidates \\
\toprule
Documents               &    --- & 4,693 \\
Annotated Documents     &  4,381 & 2,114 \\
Have Country Data       &  4,160 & 2,008 \\
\midrule
Advanced Econ Papers    &  3,374 & 1,609 \\
Developing Econ Papers  &    543 &   272 \\
Mixed Econ Papers       &    243 &   127 \\
\midrule
Advanced Econ Authors   & 10,679 & 5,250 \\
Developing Econ Authors &  2,317 & 1,187 \\
\bottomrule
\end{tabular}
    \caption{Annotation outcome summary.}
    \label{tab:outcomes}
\end{table}

Table~\ref{tab:outcomes} shows a summary of the collected annotation data, after merging and integrating data sources.  For these statistics, to aggregate each paper's authors into a single economic designation for the paper, we considered a paper to be from an advanced or developing economy if all authors' locations had the same economic designation; otherwise, we list it as a `mixed' economy paper.

\subsection{Group Definitions}
\paragraph{Group definition accompanying the training data.}
To help participants get started, we provided a file containing group membership definitions for authors in the S2
corpus. This definition was based on author h-indices.  This definition was not used in the final evaluation, but was meant as a starting point for system development. For each author, the data consisted of:
\begin{itemize}
\itemsep-0.1em 
    \item the author's S2 ID,
    \item the author's group identifier.
\end{itemize}
Authors were split into $2$ groups, based on the value of their h-index.

\paragraph{Group definitions for evaluation.}
Our primary evaluation was based on the NIST assessors' country annotations. We combined these annotations with economic development levels from the International Monetary Fund. With this definition, the fairness target is to ensure fair exposure for papers written in countries with more- and less-developed economies. The evaluation itself uses individual author-level annotations; the exposure a mixed-economy paper receives counts towards both developing and advanced economy exposure. Under this definition, authors are split into two groups.
\section{Results}
\label{sec:results}

%
%
\begin{table}[]
    \centering
    \begin{tabular}{lr}
    run & $\frEE$\\
    \hline
    NLE\_META\_9\_1	&	0.428	\\
    NLE\_META\_99\_1	&	0.429	\\
    NLE\_META\_PKL	&	0.433	\\
    NLE\_TEXT\_9\_1	&	0.438	\\
    NLE\_TEXT\_99\_1	&	0.442	\\
    UoGTrBComFu	&	0.475	\\
    LM-rel-groups	&	0.580	\\
    LM-relevance	&	0.601	\\
    MacEwan-base	&	0.722	\\
    UoGTrComRel	&	0.798	\\
    LM-relev-year	&	0.811	\\
    UoGTrBComRel	&	0.832	\\
    MacEwan-norm	&	0.850	\\
    UoGTrBComPro	&	0.851	\\
    UW\_bm25	&	0.875	\\
    UoGTrBRel	&	0.886	\\
    UW\_Kr\_r60g20c20	&	0.895	\\
    umd\_relfair\_ltr	&	0.907	\\
    UW\_Kr\_r25g25c50	&	0.916	\\
    UW\_Kr\_r0g0c100	&	0.948	\\
    UW\_Kr\_r0g100c0	&	0.999	\\
    LM-rel-year-100	&	1.046	\\
    Deltr-gammas	&	1.067	\\
    \end{tabular}
    \caption{Reranking results.  Runs ordered in increasing expected exposure (Equation \ref{eq:eemetric}).  \textit{Smaller values are better}.}
    \label{tab:reranking}
\end{table}

%
%
\begin{table}[]
    \centering
    \begin{tabular}{lr}
    run & $\frEE$\\
    \hline
    UW\_t\_bm25	&	0.748	\\
    UW\_Kt\_r80g10c10	&	0.769	\\
    UW\_Kt\_r60g20c20	&	0.770	\\
    UW\_Kt\_r25g25c50	&	0.821	\\
    UW\_Kt\_r0g0c100	&	1.056	\\
    \end{tabular}
    \caption{Retrieval results.  Runs ordered in increasing expected exposure (Equation \ref{eq:eemetric}). \textit{Smaller values are better}.}
    \label{tab:retrieval}
\end{table}

\begin{figure}
	\begin{center}
	\includegraphics[width=5in]{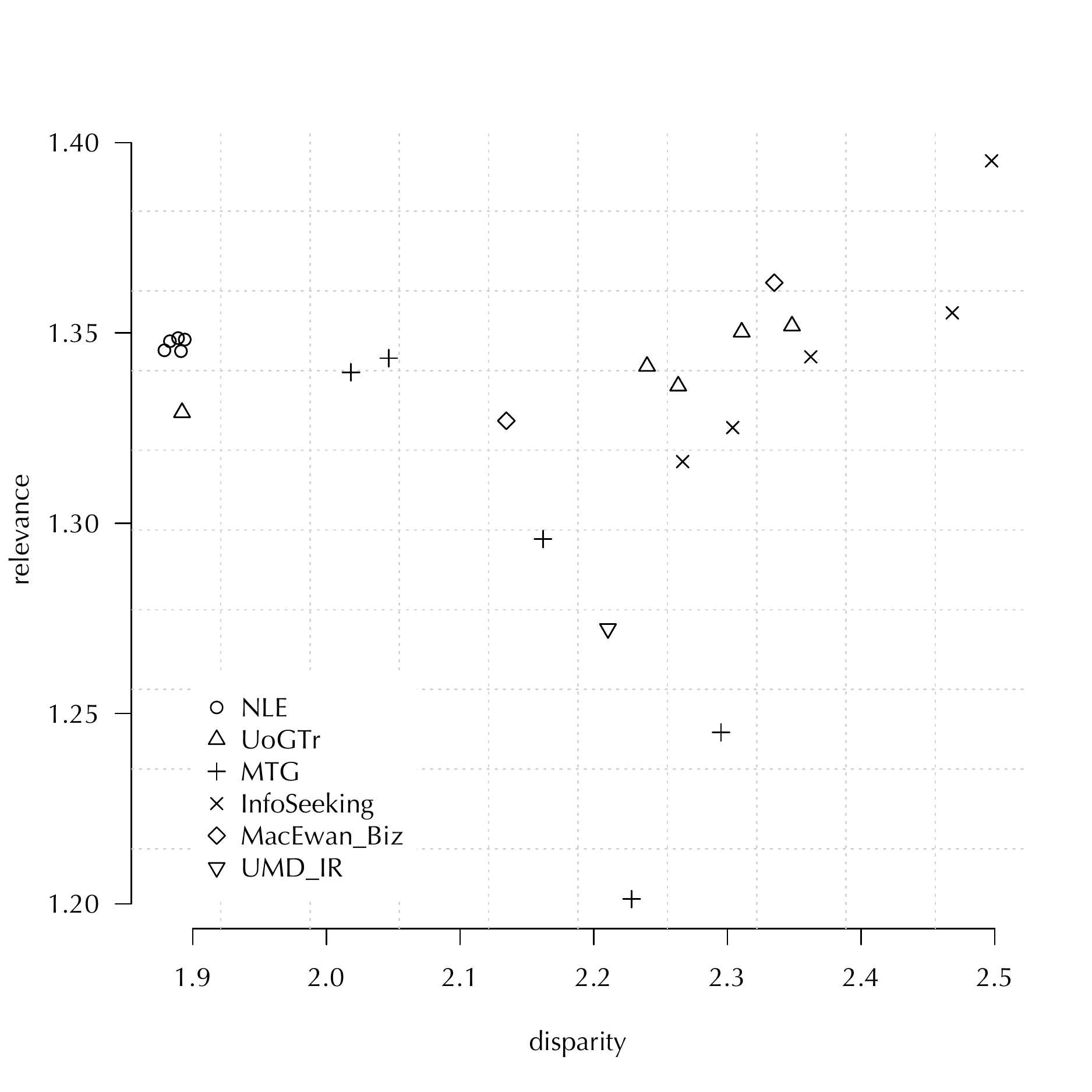}
	\end{center}
	\caption{Disparity and relevance results for the reranking task.  Lower disparity values  are better.  Higher relevance values are better.}
	\label{fig:DR}
\end{figure}

\subsection{Submitted runs}

This year, 6 different teams submitted a total of 28 runs, (23 runs for the reranking task and 5 runs for the retrieval task).
Some of the approaches included: 
\begin{itemize}
    \item weighted reranking methods that optimized the KL-divergence of the output group distributions to target group distributions estimated using external Google Scholar data (team InfoSeeking),
    \item ranking fusion methods where the documents were ranked by the BM25 scores of queries matched to different document parts (title, abstract) and the individual ranking weights shift throughout a query sequence (team MacEwan),
    \item including authors from all groups at the top of the ranking with the groups determined by a clustering algorithm on the authorship graph; an approach that randomizes the output of a learning-to-rank algorithm based on the predicted relevance and optionally includes the publication year as a feature (team MTG),
    \item randomization of the outputs of rankings based on the textual content of documents and externally trained word embeddings, with an optional parametrized readjustments to match a target group exposure distribution (team NLE),
    \item a static method that does keep track of exposure in between rankings based on a learning-to-rank algorithm with a custom objective balancing fairness and relevance (team UMD),
    \item a two-stage approach where the first stage is based on standard retrieval methods, and the second stage uses reranking based on membership in authorship communities detected using graph embedding methods (team UoGTr).
\end{itemize}

Notably, novel ideas as compared to the last year's runs included estimating group membership using automatically detected authorship communities, randomization of the outputs, including the publication year as a feature, and incorporation of external resources (Google Scholar data and word embeddings trained on the Bing corpus).

\subsection{Evaluation}
We present the results for reranking and retrieval in Tables \ref{tab:reranking} and \ref{tab:retrieval}, sorted by $\frEE$.  Although the run descriptions were not sufficient to draw many general conclusions, the top runs all used external resources (e.g. public embeddings) and multiple permutations per query (e.g. amortization or randomization).  

We can decompose $\frEE$ into relevance and disparity components \cite{diaz:expexp},
\begin{align}
    \text{disparity} &= \sum_{\frGroup\in\frGroups}\frGroupExposure_\frGroup^2 \\
    \text{relevance} &= \sum_{\frGroup\in\frGroups}\frGroupExposure_\frGroup\times \frGroupExposure_\frGroup^* 
\end{align}
This allows us to plot each run on disparity-relevance axes which often reflects a trade-off between disparity and relevance.  We present results in Figure \ref{fig:DR}.  In general, we would like runs to lie close to the top left corner.  Although the top performing runs from NLE had relatively high relevance, the strong $\frEE$ was more attributable to exhibiting less disparity.

\bibliography{notebook} 
\end{document}